\documentclass[%
 reprint, 
 amsmath,amssymb,
 aps,
]{revtex4-2}
    
\usepackage{graphicx}
\usepackage{dcolumn}
\usepackage{bm}
\usepackage{amsmath}
\usepackage{amssymb}
\usepackage{txfonts}
\usepackage{mathrsfs}
\usepackage{xcolor}
\usepackage{subcaption}
\usepackage[normalem]{ulem}
\usepackage[mathlines]{lineno}

\begin{document}

\title{Low-temperature behavior of the $O(N)$ models below two dimensions}
\author{Andrzej Chlebicki}
\affiliation{Institute of Theoretical Physics, Faculty of Physics, University of Warsaw, Pasteura 5, 02-093 Warsaw, Poland}
\author{Pawel Jakubczyk }
\affiliation{Institute  of Theoretical Physics, Faculty of Physics, University of Warsaw, Pasteura 5, 02-093 Warsaw, Poland}
\date{\today}
\begin{abstract}
We investigate the critical behavior and the nature of the low-temperature phase of the $O(N)$ models treating the number of field components $N$ and the dimension $d$ as continuous variables with a focus on the $d\leq 2$ and $N\leq 2$ quadrant of the $(d,N)$ plane. We precisely chart a region of the $(d,N)$ plane where the low-temperature phase is characterized by an algebraic correlation function decay similar to that of the Kosterlitz-Thouless phase but with a temperature-independent anomalous dimension $\eta$. We revisit the Cardy-Hamber analysis leading to a prediction concerning the nonanalytic behavior of the $O(N)$ models' critical exponents and emphasize the previously not broadly appreciated consequences of this approach in $d<2$. In particular, we discuss how this framework leads to destabilization of the long-range order in favour of the quasi long-range order in systems with $d<2$ and $N<2$. Subsequently, within a scheme of the nonperturbative renormalization group we identify the low-temperature fixed points controlling the quasi long-range ordered phase and demonstrate a collision between the critical and the low-temperature fixed points upon approaching the lower critical dimension. We evaluate the critical exponents $\eta(d,N)$ and $\nu^{-1}(d,N)$ and demonstrate a very good agreement between the predictions of the Cardy-Hamber type analysis and the nonperturbative renormalization group in $d<2$.
\end{abstract}

\maketitle

\section{Introduction}
The $O(N)$ models are a cornerstone of modern statistical physics. They have been employed with a great success to describe phase transitions of a multitude of universality classes: the liquid-vapor transition for $N=1$, the helium superfluid transition for $N=2$, the polymer problem for $N=0$ and the scalar sector of the electroweak standard model for $N=4$ to invoke just a few. Additionally, their relatively simple formulation has provided a fertile ground for development and evaluation of new theoretical techniques. 

The universality classes of the $O(N)$ models [and the associated critical exponents] can be parameterized by two variables: $d$ - the system's spatial dimension and $N$. Typically, only the integer pairs of $d$ and $N$ are of interest from a physical standpoint. However, there is a number of significant theoretical approaches treating both $d$ and $N$ as continuous variables. The most notable examples include: the large $N$ expansion, the $4-\epsilon$ expansion and the $2+\epsilon$ expansion. These perturbative renormalization group (RG) approaches have proven successful both in predicting the values of the critical exponents as well as development of the general theory of RG \cite{Amit2005,Zinn-Justin2010}. We note, that there has been some effort in engineering systems with non-integer effective values of $d$ \cite{Boada2012,Boada2015,ebek2020} and that recently a rigorous  mathematical description has been developed for systems with non-integer $N$ \cite{Binder2020}. The idea of considering non-integer values of $d$ and $N$ was also implemented in the conformal bootstrap approach [see e.g. \cite{Showk_2014, Henriksson_2022}].

Depending on $d$ and $N$, the $O(N)$ models can support three very distinct types of phases: long-range ordered (LRO), disordered and quasi long-range ordered (QLRO). A LRO phase exhibits a non-vanishing order-parameter expectation value. A disordered phase, on the other hand, is characterized by vanishing order-parameter and additionally an exponential decay of the correlation function $G(r) \propto \exp(-r/\xi)$, where $\xi$ denotes the correlation length. A QLRO phase, alike a disordered phase, is characterized by vanishing order-parameter, but the corresponding correlation function decays as a power-law of the distance $G(r) \propto r^{2-d-\eta}$, where $\eta$ is the anomalous dimension. This correlation structure arises when the RG flow is controlled by a stable finite-temperature fixed point (FP). 

\begin{figure}
    \centering
    \includegraphics[width=.45 \textwidth]{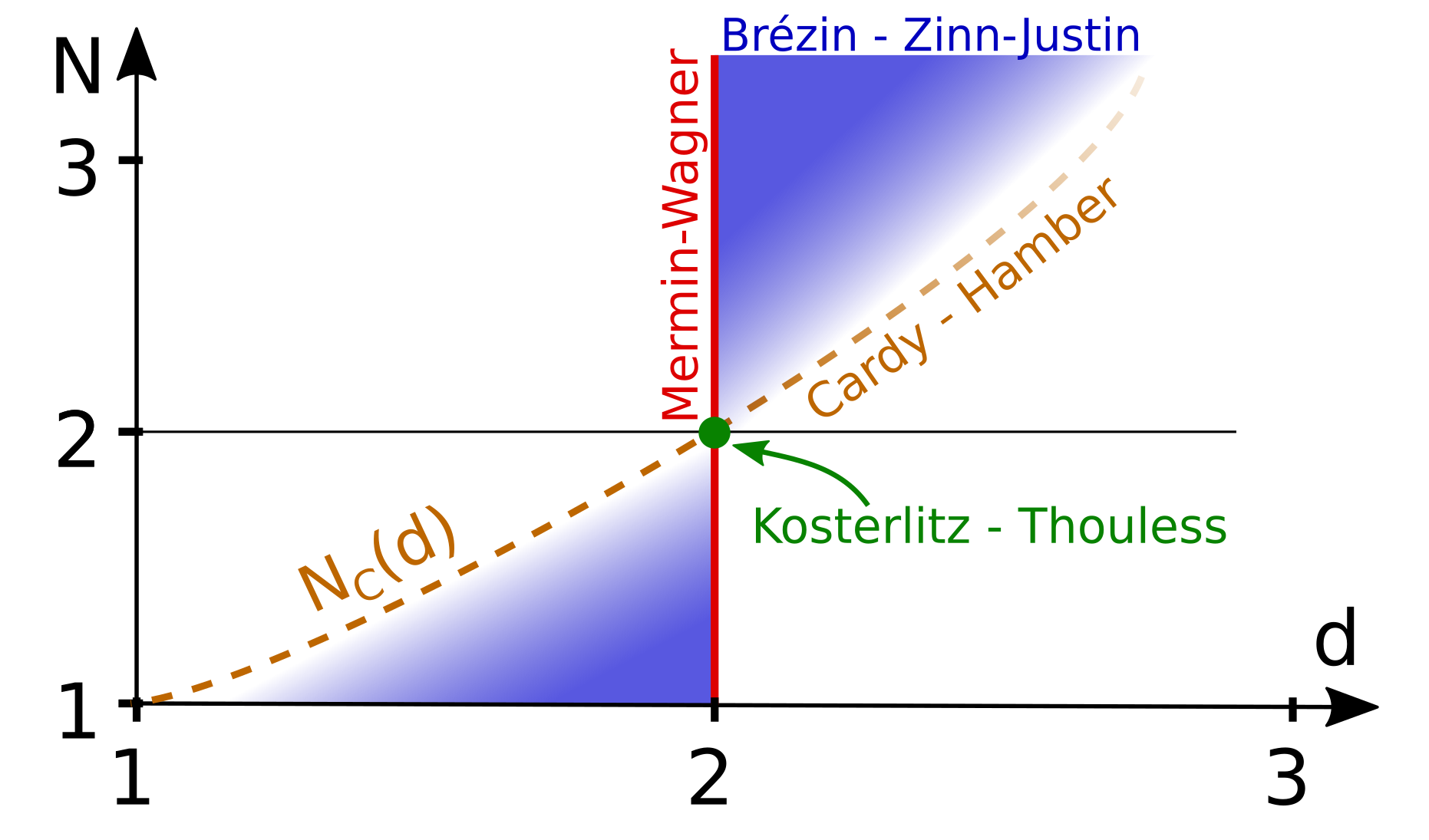}
    \caption{(Color online) Schematic representation of the $(d,N)$ plane in the vicinity of the Kosterlitz-Thouless point $(2,2)$. The Mermin-Wagner line [$d=2$, $N>1$] separates the systems that support the long-range order (LRO) in non-zero temperatures from those that do not. Below two dimensions, for sufficiently small values of $N$ [$N<N_c(d)$] the low-temperature phase is characterized by QLRO, while for $N>N_c(d)$ the system remains disordered for any non-zero temperature. For $d<2$, the Cardy-Hamber line coincides with $N_c(d)$, while for $d>2$ it is a locus of the hypothetical nonanalyticity of the critical exponents predicted within the framework of ref. \cite{Cardy1980}, but contrary to the findings of ref. \cite{Chlebicki2021}.}
    \label{fig:main_diagram}
\end{figure}

Over the years the unusual physics of QLRO phases attracted significant attention. The most famous example is probably the low-temperature phase of the Kosterlitz-Thouless universality class $(d,N)=(2,2)$ \cite{Kosterlitz_1973}. This exotic type of ordering was also found, \emph{inter alia}, in the random-field [or random-anisotropy] $O(N)$ models - relevant in the studies of superconductors, nematic liquid crystals and porous media \cite{Feldman2001, Tarjus2020}. For specific values of $d$ and $N$, a weak random-field constitutes a relevant perturbation which destroys the LRO but allows for the formation of the QLRO sometimes called the Bragg glass phase \cite{Giamarchi1994, Feldman2001}. The universal properties of the random-field $O(N)$ models are related to those of the pure $O(N)$ models in a shifted dimension due to a property known as dimensional reduction \cite{Parisi1979, Tissier2006, Tarjus2020}. Fully grasping the critical behavior of the $O(N)$ models is therefore relevant for a wide variety of physical contexts. Surprisingly, an old finding that the $O(N)$ models support a QLRO phase in an extended region of the $(d,N)$ plane seems relatively unknown and poorly studied \cite{Lau1987,Tissier2006}.

A sketch of the $(d,N)$ plane is presented in fig. \ref{fig:main_diagram}. The point $(2,2)$ marks a unique spot in the $(d,N)$ plane, which hosts the exotic Kosterlitz-Thouless (KT) universality class. The line $d=2$ marks the onset of the $2+\epsilon$ expansion. The Mermin-Wagner theorem states that at finite-temperature spontaneous breaking of the continuous symmetry can only occur above two dimensions \cite{Mermin1966}. Therefore, the $d=2$ Mermin-Wagner line separates the systems that support LRO in finite temperatures from those that do not. The meaning of the `continuity' of the symmetry group becomes unclear when we consider non-integer values of $N<2$, however based on the results from earlier studies \cite{Lau1987, Tissier2006} as well as the calculations of the present paper we argue that the Mermin-Wagner line extends all the way down to $N=1$ [not including the point $(d,N) = (2,1)$].

The KT theory perturbatively captures the way in which vortices renormalize the spin-wave stiffness, but remains restricted to a single point in the $(d,N)$ plane. The $2+\epsilon$ expansion, on the other hand, strictly describes the spin-wave theory in the vicinity of the KT point $(2,2)$, but neglects the effects of the topological excitations, which are relevant for some phase transitions, e.g. at the points $(2,2)$ and $(3,3)$ [see refs. \cite{Kamal1993, Motrunich2004} ]. It is natural to ask, how these two approaches can be combined. This question was first posed by Cardy and Hamber \cite{Cardy1980}. By combining the RG equations of KT and the $2+\epsilon$ expansion they predict a collision of fixed points upon crossing a specific line in the $(d,N)$ plane with important consequences for the model's universal properties. 

In the present study, we first revisit the Cardy-Hamber (CH) scenario with a focus directed towards the bottom-left quadrant of the $(d,N)$ plane [$d<2$, $N<2$]. The complementary case of $d\geq 2$ was addressed in our earlier paper [see ref. \cite{Chlebicki2021}]. We begin by summarizing the original CH theory. We then explain how the CH approach predicts the appearance of the QLRO phase in a specific region of the $(d,N)$ plane - the conclusion not mentioned in the CH study, however recognized in Ref.~\cite{Tissier2006}. We subsequently confront these results with a direct numerical evaluation of the nonperturbative renormalization group equations. In this framework we calculate the shape of the line $N_c(d)$ bounding the region where the QLRO low-temperature phase exists in the system's phase diagram [see fig. \ref{fig:main_diagram}]. In the vicinity of the KT point, we find this line laying very close to the CH prediction. We, finally, calculate the anomalous dimension $\eta$ of the critical and low-temperature fixed points and the correlation length exponent $\nu$ as a function of $d$ and $N$.

The present paper has the following structure. In section II we readdress the CH scenario in detail and lay out its previously not  appreciated (except Ref.~\cite{Tissier2006}) consequences. In particular, we show how the fixed point of the $2+\epsilon$ expansion becomes infrared stable for $d<2$ and $N<2$, leading to formation of the QLRO in low temperatures. Somewhat surprisingly this fact is not discussed in the CH paper. In section III we introduce the framework of the nonperturbative renormalization group and explain the employed approximations. Section IV contains our results for the shape of the line $N_c(d)$ as well as the critical exponents of the $O(N)$ models below two dimensions. In section V we summarize our conclusions.

\section{Cardy-Hamber approach}
The approach proposed by Cardy and Hamber in \cite{Cardy1980} relies on the assumption that the RG equations of the KT theory \cite{Jose1977} and the $2+\epsilon$ expansion \cite{Brezin1976b} can be analytically combined. This way we can create a single set of recursion relations for two variables: $g$ - the spinwave interaction coupling and $y^2$ which for $N=d=2$ describes the vortex fugacity and has a somewhat ambiguous physical interpretation for other values of $d$ and $N$. In specific contexts parameters analogous to $y^2$ may retain their physical meaning in different dimensions \cite{Herbut_1998, Herbut_2000}.

When investigating phase transitions close to the KT point, we treat $\epsilon=d-2$, $N-2$ and $y^2$ as small quantities of the order no higher than $O(\epsilon)$. This leads \cite{Cardy1980} to the equations:
\begin{align}
    &\dot{g} = -\epsilon g + (N-2) f(g, N) + 4\pi^3 y^2 + O(\epsilon^2),\label{ch_eqs}\\
    &\dot{y^2} = \left(4-\frac{2\pi}{g}\right) y^2 + O(\epsilon^2),\nonumber
\end{align}
where $f(g, N)$ is a function which so far has only been studied perturbatively in powers of $g$. The following reasoning is fairly general and does not require the knowledge of the exact shape of $f(g,N)$, only the assumption that $f(g,2)/g$ is a monotonic increasing function for small values of $g$. However, later on, we are going to exploit the expansion of $f$ to the highest known order \cite{Bernreuther1986}:
\begin{align}
    f(g, N) &= \frac{g^2}{2 \pi } + \frac{g^3}{4 \pi ^2}+\frac{g^4 (N+2)}{32 \pi ^3} \label{f_perturbative}\\
    &-\frac{g^5 \left(N^2-18 N \zeta (3)-22 N+54 \zeta (3)+34\right)}{192 \pi ^4}+O(g^6).\nonumber
\end{align}

In the analysis of eqs. \eqref{ch_eqs} an essential role is played by the parameter
\begin{equation}
    \Delta = \epsilon \frac{\pi}{2} - (N-2) f\left(\frac{\pi}{2},N\right) +O(\epsilon^2),\label{delta}
\end{equation}
the sign of which determines the existence and stability of the fixed point solutions to eq. \eqref{ch_eqs}. The condition $\Delta=0$ defines the CH line at which the distinct families of fixed point solutions collide. As we shall see, for $\epsilon<0$ this line marks also the lower critical dimension.

Equations \eqref{ch_eqs}, admit two families of non-trivial fixed point solutions:
\begin{align}
    &g_{\text{KT}} = \frac{\pi}{2} + O(\epsilon), \quad y_{\text{KT}}^2 = \frac{\Delta}{4\pi^3},
\end{align}
which can be seen as an extension of the Kosterlitz-Thouless FP to $N\neq 2$, $d \neq 2$, and
\begin{align}
    \epsilon g_{\text{BZJ}} = (N-2) f(g_{\text{BZJ}},N) + O(\epsilon^2), \quad y_{\text{BZJ}}^2 = O(\epsilon^2) 
    \label{bzj_fp}
\end{align}
identical to the solution of the $2+\epsilon$ expansion of Br\'ezin and Zinn-Justin \cite{Polyakov1975, Brezin1976b}. The KT FP solution is physical [$y \in \mathbb{R}$] whenever $\Delta \geq 0$, i.e. below the CH line [see fig. \ref{fig:main_diagram}]. The analysis of the RG eigenvalues shows that this FP, if it exists, has a single unstable direction and therefore controls the phase transition.

Assuming the form \eqref{f_perturbative} of $f$, the condition \eqref{bzj_fp} can be satisfied by a number of nontrivial solutions, of which all but one can be discarded as nonphysical artifacts of the $g$ expansion. From now on, we shall use the term BZJ FP to describe the solution to $\eqref{bzj_fp}$ with the lowest positive value of $g_{\text{BZJ}}$, which is the only physical. The BZJ FP exists in the top-right [$d>2$, $N>2$] and bottom-left [$d<2$, $N<2$] quadrants of the $(d,N)$ plane. In the top-right quadrant, the BZJ FP is critical for $\Delta<0$ and tricritical for $\Delta>0$. For $\Delta=0$, BZJ and KT FPs coincide and swap roles, which leads to the hypothetical nonanalyticity of the critical exponents (predicted by this framework). We discuss this part of the CH scenario in ref. \cite{Chlebicki2021}.

A different picture presents itself in the bottom-left quadrant of the $(d,N)$ plane [see fig. \ref{fig:main_diagram}], which is the main focus of the present study. The CH paper notes that the KT FP is critical and controls the phase transition when $\Delta>0$ and that there is no phase transition for $\Delta<0$. However, Cardy and Hamber also state that only the KT FP is real for $N<2$. It is nonetheless  rather evident from eq.~\eqref{ch_eqs}, the BZJ FP also exists in this quadrant, and, as we argue below, controls the  low-temperature phase present in the $\Delta>0$ part of the $(d,N)$ plane below $d=2$. This fact was first observed in Ref.~\cite{Tissier2006}

For negative values of $\epsilon$, the zero-temperature trivial fixed point [$g=0$] is unstable, which prevents formation of the long-range order at any finite temperatures - in accord with the Mermin-Wagner theorem. However, at the same time, a new infrared stable FP appears. Namely, for $\epsilon<0$ and $\Delta>0$, the BZJ FP is a stable finite-temperature FP which emerges from the zero-temperature FP as $\epsilon$ crosses $0$. In this region, the KT FP controls the transition between the disordered and the QLRO phases. 

Above the CH line [see fig. \ref{fig:main_diagram}], when $\Delta$ becomes negative, the KT FP ceases to exist. At the same time, the BZJ FP attains a single relevant direction while its attraction domain shrinks to a subset of the $y^2=0$ line. This fixed point does not control any phase transition. This is because, as we deduce from eq. \eqref{ch_eqs}, the RG flows cannot cross between the regions $y^2=0$ and $y^2>0$. These two regions differ with respect to the presence of the vortex-like excitations, which can appear when $y^2>0$, but not when $y^2=0$. For this reason, the systems corresponding to those two regions are subject to very distinct physics. The present study primarily focuses on the models with $y^2>0$. From this perspective BZJ FP acts simply as a repulsive FP for $\Delta<0$. We, therefore, expect that no phase transition takes place for $\Delta<0$ and consequently that the condition $\Delta=0$ defines the line of lower critical dimensions [in the bottom-left quadrant of the $(d,N)$ plane].

\begin{figure}
    \centering
    \includegraphics[width=.45 \textwidth]{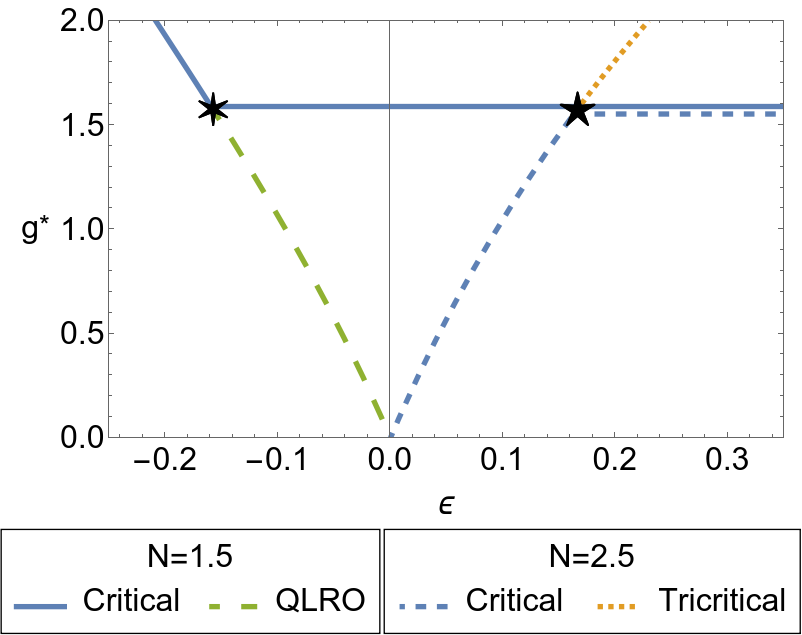}
    \caption{(Color online) Illustration of the Cardy-Hamber scenario for the fixed point collision. The figure presents the evolution of different fixed point solutions $g^*(\epsilon,N)$ upon varying $\epsilon$ for $N=1.5$ and $N=2.5$. The six-pointed star denotes a point of FP collision for $N=1.5$, the five-pointed - for $N=2.5$. The horizontal lines mark the overlapping critical fixed points $g^*(\epsilon, 2.5)=g^*(\epsilon,1.5)=g_{\text{KT}}=\pi/2$ and were slightly separated for the purpose of transparency.}
    \label{fig:cardy_sol}
\end{figure}

The CH scenario for the fixed point collision is illustrated in fig. \ref{fig:cardy_sol}. We emphasize the contrast between the physical implications of the CH scenario above and below $d=2$. For $d>2$, the collision takes place between the critical and the tricritical FPs. The second order phase transition takes place between the disordered and the LRO phases, and the critical exponents exhibit (within this approach) nonanalyticities at the line of the collisions [$\Delta=0$]. For $d<2$, on the other hand, the critical FP collides with the FP controlling the low-temperature behavior. The phase transitions take place between the disordered and the QLRO phases, and the line of the collisions marks the lower critical dimensions $d_c(N)$. The above picture, as obtained for $d<2$, is confirmed with the nonperturbative renormalization group (NPRG) calculations described in sec. \ref{sec:NPRG}. Observe, that this remains in contrast to the case $d>2$ addressed in Ref.~\cite{Chlebicki2021}.

\section{Nonperturbative renormalization group}
\label{sec:NPRG}
To readdress the presented problem from a different point of view, we employ the one-particle irreducible variant of the nonperturbative renormalization group \cite{wetterich}. The central object in the NRPG is a scale-dependent functional $\Gamma_k$ of the order-parameter $\phi$, called the effective average action. The quantity $\Gamma_k[\phi]$ describes the system's effective free energy at scale $k$, treating the slow [$q<k$] order-parameter modes $\phi_q$ at the mean-field level while including the fluctuations of the fast modes [$q>k$]. In this way, it smoothly interpolates from the mean-field free energy at the ultraviolet scale $k=\Lambda$ to the exact free energy at the vanishing scale $k\rightarrow 0$. The RG flow of the effective action is described by the Wetterich equation \cite{wetterich}:
\begin{equation}
    \partial_k \Gamma_k = \frac{1}{2}\mathrm{Tr}\left[ \partial_k R_k \left(\Gamma_k^{(2)}+ R_k\right)^{-1}\right], \label{wetterich}
\end{equation}
where $\Gamma_k^{(2)}$ denotes the second functional derivative of $\Gamma$ and $R_k$ is the infrared regulator function. The NPRG has been applied with remarkable success in a wide variety of contexts ranging from statistical mechanics and quantum many-body physics to gravity and high-energy physics [for reviews see e.g. \cite{berges, Kopietz2010, Gies2012, Dupuis2020} ]. 

\subsection{Derivative expansion}
The Wetterich equation \eqref{wetterich} is not directly soluble. We employ the derivative expansion (DE) to reduce the functional differential equation \eqref{wetterich} to a set of partial differential equations. The principal idea of this approximation relies on the expansion of the effective action $\Gamma_k$ around the uniform field configuration in powers of gradient operators acting on the order-parameter to a specified order. In the present work, we adopt the $O(\nabla^2)$ truncation, with $\Gamma_k$ parameterized as
\begin{equation}
    \Gamma_k = \int d^dx \left\{U^k(\rho) + \frac{Z_2^k (\rho)}{2} \left(\nabla\phi\right)^2 + \frac{\left(Z_1^k(\rho)-Z_2^k(\rho)\right)}{4\rho}\left(\nabla\rho\right)^2\right\}, \label{ansatz}
\end{equation}
including all $O(N)$-symmetric terms up to $\nabla^2$. In the effective action ansatz, the invariant $\rho = \phi^i \phi^i/2$ was introduced to make the $O(N)$ symmetry of $\Gamma_k$ manifest and simplify the RG flow equations. Note that no field expansion is imposed here.

Having specified the ansatz, the flow equations are derived in a straightforward but tedious calculation. The function $\beta^k_U(\rho)=k\partial_k U^k(\rho)$ is obtained by plugging the ansatz \eqref{ansatz} into \eqref{wetterich} and evaluating at a uniform field configuration $\phi^1(x)^2 = 2 \rho$, $\phi^i(x)=0$ for $1<i\leq N$. The derivation of $\beta^k_{Z_1}$ and $\beta^k_{Z_2}$, defined as $\beta^k_{Z_i}(\rho) = k\partial_k Z^k_i(\rho)$ first requires us to differentiate eq. \eqref{wetterich} functionally and subsequently expand in momenta:
\begin{equation}
    \beta_{Z_{i}} = \left.\frac{1}{2 d}\left.\Delta_p\right|_{p=0} \frac{\delta^2\left( \partial_k \Gamma_k\right)}{\delta \phi^{i}_p \delta \phi^{i}_{-p} }\right|_{\mathrm{Unif.}}.
\end{equation}
Due to the significant length of the flow equations, we relegate their presentation to the Appendix. The flow equations are fully equivalent to those analyzed in ref. \cite{Chlebicki2021}.

To make the fixed point behavior manifest, we follow a standard rescaling procedure to express the parameterization in a dimensionless form. This is achieved by multiplying each quantity $X$ by the scale raised to the power of its canonical dimension $\tilde X = X k^{d_x}$. The rescaled variables read:
\begin{equation}
    \tilde \rho = \rho k^{2-d -\eta_k}, \quad \tilde U^t(\tilde \rho) = U^k(\rho) k^{-d}, \quad \tilde{Z}^t_{1/2}(\tilde \rho) = Z^k_{1/2}(\rho) k^{\eta_k}.
\end{equation}
Above, $\eta_k$ denotes the running anomalous dimension defined as a logarithmic derivative of the scaling factor $\eta_k = - k\partial_k \log\left(Z^k\right)$. We close the set of RG equations by defining the rescaling factor $Z_2(\rho_\eta)/Z_k = \tilde Z_2(\tilde \rho_\eta) \equiv 1$, with an arbitrary constant $\tilde \rho_\eta$. The equations are further transformed into a stationary form by introducing the `renormalization time' $t=\log(k/\Lambda)$ [$\partial_t = k\partial_k$].

With the flow equations derived, we follow a two-pronged approach to the described problem. The first avenue relies on a direct integration of the flow equations to inspect the phase diagram of the $O(N)$ models. As an initial condition of the RG flow we choose the `mexican hat'-type $O(N)$ symmetric action:
\begin{equation}
    \Gamma_\Lambda[\phi] = \mathcal{S}[\phi] = \int d^dx \left(\frac{1}{2}\left(\nabla \phi\right)^2+ \frac{u}{8}\left(\phi^2-\phi_0^2\right)^2\right),
\end{equation}
with $\phi_0$ tuned to approach the critical domain for $k\rightarrow 0$. Simultaneously, we follow a complementary path based on solving the fixed point equation
\begin{equation}
    \partial_t \tilde \Gamma^* = 0.
\end{equation}
This allows us to extract the values of the critical exponents through linearization of the flow equations and identify the lower critical dimension below which the fixed point no longer exists. In our flow equations, $d$ and $N$ enter as numerical constants. We use that fact in both these approaches to observe how the critical behavior changes as we traverse the $(d,N)$ plane.

To make the flow equations suitable for the numerical analysis, we introduce what we now call the functional scheme. In this approach, we represent the functions $\tilde V(\tilde\rho) = \tilde U'(\tilde\rho)$, $\tilde Z_1(\tilde\rho)$ and $\tilde Z_2(\tilde\rho)$ on a finite discrete grid. We approximate the derivatives by finite differences at  orders either $O(h^4)$ or $O(h^8)$ depending on the calculation, the momentum integrals by finite sums using the Gauss-Legendre quadrature and employ the adaptive $4^{\text{th}}$ order Runge-Kutta-Fehlberg algorithm when integrating the flow equations. This way we recast the derived set of partial differential equations into a very large set of algebraic equations that can be solved numerically. We thoroughly investigated the dependence of our results on the numerical parameters of the discrete representation. We used a $121$ point $\tilde \rho$-space grid to obtain the presented numerically converged results.

Throughout the work, we adopt the Wetterich regulator function \cite{berges}:
\begin{equation}
    R_k(q^2) = \alpha Z_k k^2 \frac{ \frac{q^2}{k^2}}{\exp\left(\frac{q^2}{k^2}\right)-1},
\end{equation}
with a variable parameter $\alpha$. In accord with the principle of minimal sensitivity (PMS) \cite{Balog2020, Canet2003, DePolsi2020}, our estimate of a physical quantity corresponds to the value of $\alpha$ such that the quantity in question is stationary with respect to $\alpha$. 

\subsection{Low-temperature behavior}
\label{lowt_section}
When investigating the low-temperature behavior, unlike in the case of the critical state or the high-temperature behavior, the functional approach presented above is notoriously difficult to implement. The problem lies in the structure of the flow equations, which can be depicted as one-loop diagrams with lines representing `dressed' propagators of the form [scale dependence suppressed in the notation for clarity]:
\begin{align}
    &G_1^{-1}(\rho, q) = U'(\rho) + 2 \rho U''(\rho) + q^2 Z_1(\rho) + R(q^2), \\
    &G_2^{-1}(\rho, q) = U'(\rho) + q^2 Z_2(\rho) + R(q^2). \nonumber
\end{align}
Both propagators and consequently each term in the $\beta$ functions, suffer from poles in $q$. At finite scales these poles remain on the imaginary axis, however as we approach the infrared limit in the low-temperature phase, $U'(0) + R(0) \rightarrow 0^+$ taking the poles closer to the real axis and leading to numerical inaccuracies in the momentum integrals' evaluation.

To a certain extent, this problem can be dealt with by adopting very precise procedures for carrying out the momentum integrals. However, the numerical cost of achieving a given precision diverges rapidly as the pole of the propagator approaches the real axis. The precision achieved in our numerical implementation allowed us to capture the low-temperature fixed point inside a region roughly demarcated by the CH line from the top and the lines $N=1.25$ and $d=1.9$ from the bottom and the right respectively [compare fig. \ref{fig:main_diagram}]. We were unable to numerically push the bounds of the region closer to the Mermin-Wagner line ($d=2$) or the Ising universality class ($N=1$). This is explained by the fact that the low-temperature behaviour of the models lying on these two lines is controlled by singular zero-temperature FPs. In the proximity of these lines, the low-temperature FPs are no longer singular, yet approaching them, though in principle possible, becomes extremely challenging from the numerical perspective. 

An alternative route for avoiding these problems relies on the expansion of the parametrizing functions into power-series around the minimum of the local potential. In the most common approach, the functions $Z_1(\rho)$ and $Z_2(\rho)$ are reduced to flowing constants while the local potential is expanded to a given order - typically such that the quantities of interest converge to specified precision. This procedure has proven valuable for studying models with $d\gtrsim 2.5$ as it converges fairly rapidly and greatly decreases the numerical complexity of the problem. It turns out, however, that in the case of $O(N)$ models with $N>1$ below dimension $d \lesssim 2.4$, this type of expansion does not reproduce the expected critical behavior when carried out beyond the lowest sensible order:
\begin{equation}
    U(\rho) \approx \frac{u_2}{2} \left(\rho-\rho_0\right)^2.\label{phi4}
\end{equation} 
The reasons for the inadequacy of the high-order field expansion in low $d$ remain unclear to us and require future clarifying studies. Henceforth, we shall refer to the scheme relying on the field expansion as the simplified scheme [in contrast to the functional scheme described earlier]. The simplified scheme will serve as a qualitatively accurate tool to outline the expected results and to describe the parts of the $(d,N)$ plane which we were not able to access within the functional scheme.

\section{Results}
\begin{figure}
    \centering
    \includegraphics[width=.45 \textwidth]{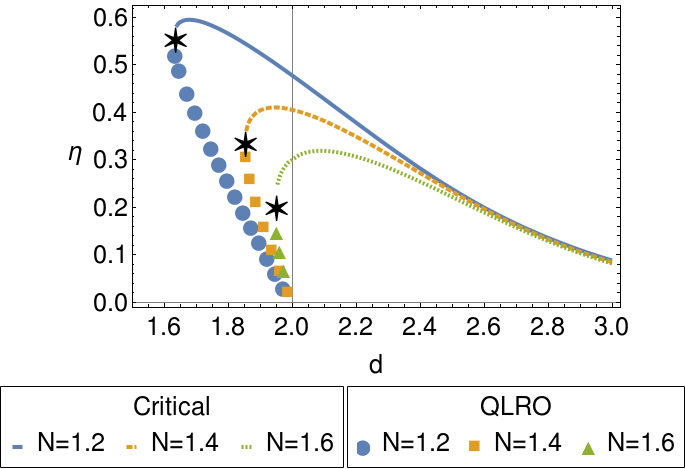}
    \caption{(Color online) Dependence of the anomalous dimension of both the critical and the QLRO fixed points on $d$ for a series of values of $N$ within the simplified scheme. Points denote the anomalous dimension of the QLRO FP and lines - the critical FP. The stars mark the collision of the FPs at the lower critical dimension as obtained by the simplified calculation.}
    \label{fig:eta_joint}
\end{figure}

Our NPRG results firmly support the CH scenario for the fixed point collision below $d=2$. This is clearly illustrated in fig. \ref{fig:eta_joint} showing the evolution of the anomalous dimensions of the critical and QLRO FPs upon decreasing the dimension calculated within the simplified scheme. The anomalous dimensions of the two FPs rise with the decreasing spatial dimension and converge abruptly when the FPs collide, upon approaching the lower critical dimension. The rest of the present section provides a detailed description of this phenomenon from the NPRG perspective with a direct comparison to the predictions of the CH scenario.

\subsection{Critical exponents}
We calculate the exponents characterizing the critical fixed point within the functional scheme in a fashion similar to ref. \cite{Chlebicki2021}. We begin by finding a fixed point in a high dimension [$d=3$]. From there we track the evolution of the critical exponents as we slowly traverse the $(d,N)$ plane along the lines of constant $N$. In practice, we achieve this by performing small, finite steps $\delta d$ using the FP solution from the point $(d+\delta d, N)$ as an initial condition for the FP search at the point $(d,N)$.

\begin{figure}
    \centering
    \includegraphics[width=.45 \textwidth]{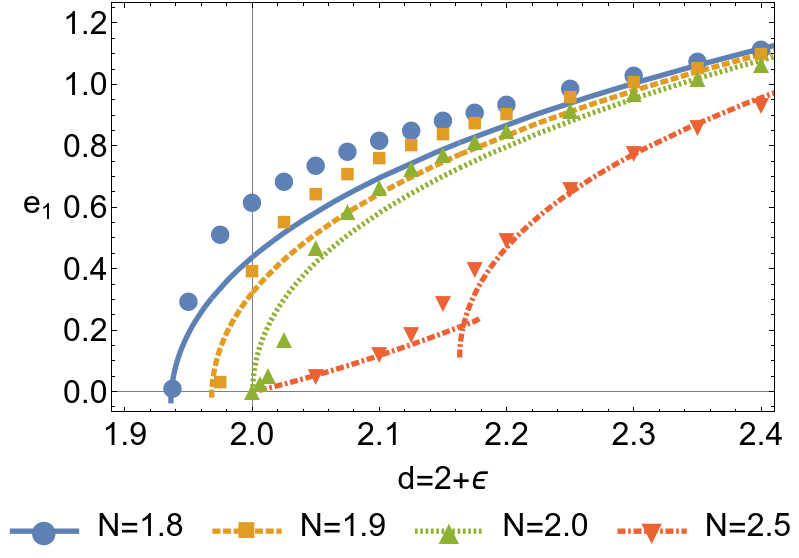}
    \caption{(Color online) Dependence of the first RG eigenvalue on $\epsilon$ for a series of values of $N$. Points denote our results from the functional calculation and lines - the predictions from the Cardy-Hamber type analysis. The red line corresponding to $N=2.5$ features a discontinuity of the first derivative predicted by the CH type analysis \cite{Cardy1980}, notably absent in the corresponding fRG calculations \cite{Chlebicki2021}. }
    \label{fig:e1_my}
\end{figure}

The results for the first RG eigenvalue $e_1=\nu^{-1}$ of the critical fixed point [representing also the inverse correlation length exponent] as a function of $d$ and $N$ are presented in fig. \ref{fig:e1_my}. The NPRG results are juxtaposed with the prediction from the Cardy-Hamber approach. Within both these approaches, all the curves behave similarly at high dimensions. However, as we lower the dimension the curve $N=2.5$ departs from the rest. Interestingly, this distinction arises almost at the same spot as the nonanalyticity predicted in the CH scenario, but in a smoothed fashion [see ref. \cite{Chlebicki2021} for an extensive discussion].

The degree to which our results for $d<2$ agree with the CH prediction is somewhat surprising taking into account the low level of expansion employed in each of the approaches. Most notably the lower critical dimension [where $e_1=0$] is almost identical in the vicinity of $d=2$. However, we also note some significant quantitative differences between the two approaches. The CH scenario predicts a square root-like shape of $e_1$ close to the lower critical dimension. This stands in contrast with our results, where $e_1$ decays with an exponent of around $0.5$ only for $N=2$, which then falls rapidly to around $0.3$ for $N=1.8$ and $0.2$ for $N=1.2$.

\begin{figure}
    \centering
    \includegraphics[width=.45 \textwidth]{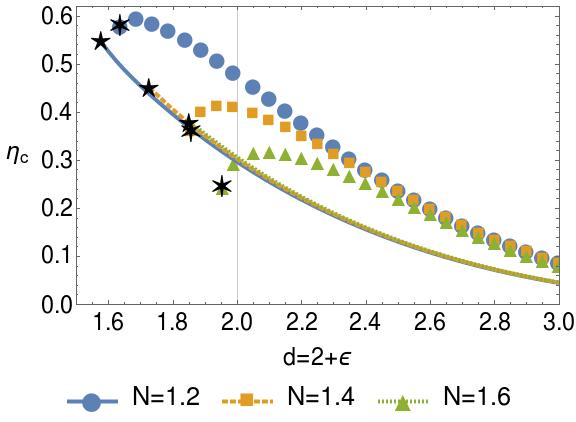}
    \caption{(Color online) Dependence of the anomalous dimension of the critical FP $\eta_c$ on $\epsilon$ for a series of values of $N$. Points denote the results from the simplified scheme and lines - the data from the functional calculation. The values at the $N$-dependent lower critical dimensions $\eta_c(d_c(N))$ are marked by the stars: five-pointed for the functional calculations and six-pointed for the simplified calculations. 
    }
    \label{fig:eta_crit}
\end{figure}

Fig. \ref{fig:eta_crit} compares the results for the anomalous dimension of the critical FP $\eta_c$ obtained within the functional scheme with those calculated within the simplified scheme. In the considered range of values, the data from the functional scheme shows very weak $N$-dependence of $\eta$ for any dimension $d$. This is, however, not the case in the simplified scheme, where $\eta$ seems weakly $N$-dependent only in high dimensions. As we lower the dimension, $\eta_c$ starts to depend on $N$ more strongly becoming non-monotonous very close to the lower critical dimension. This figure visualizes a well-established fact, that the employed form of the simplified scheme tends to significantly overestimate the anomalous dimension - in our case by a factor ranging from $3/2$ in low dimensions to even $2$ close to $d=3$. 

\subsection{Low-temperature phase}
\begin{figure}
    \centering
    \includegraphics[width=.45 \textwidth]{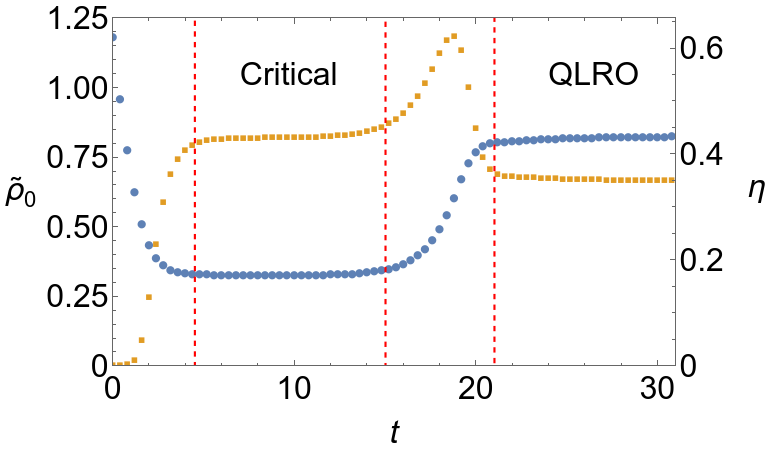}
    \caption{(Color online) Flow of the dimensionless potential minimum $\tilde \rho_0$ (left axis, blue circles) and the anomalous dimension $\eta$ (right axis, orange squares) for $(d,N) = (1.75, 1.3)$ slightly below the critical temperature calculated within the functional scheme. Red vertical lines roughly demarcate the scales at which the flow is controlled by the critical and QLRO FPs.}
    \label{fig:flow}
\end{figure}

The fixed point controlling the low-temperature phase can be extracted from the RG flow. We identify it as the limit of the effective action for vanishing scale $k\rightarrow 0$ in the flow initiated below the critical temperature. This means that the search for the low-temperature fixed point, in principle, should be easier than for the critical one, as it requires no tuning of the initial condition. However, properly capturing the low-temperature FPs within the functional scheme is notoriously difficult. One reason for that was laid out in sec. \ref{lowt_section} and is related to the pole of the propagator. The other reason lies in the abruptness of the flow in the transient regime in between the critical and the low-temperature FPs. This abruptness, somewhat surprisingly, becomes less pronounced the closer to the critical temperature we get. An example of the RG flow passing close to the critical FP and then turning to asymptotically approach the low-temperature FP is presented in fig.~\ref{fig:flow}. The figure overlays flows of two quantities: the dimensionless potential minimum $\tilde \rho_0$ and the anomalous dimension $\eta$ for $(d,N) = (1.75, 1.3)$ slightly below the critical temperature.

\begin{figure}
    \centering
    \begin{subfigure}{.48 \linewidth}
    \includegraphics[width= \textwidth]{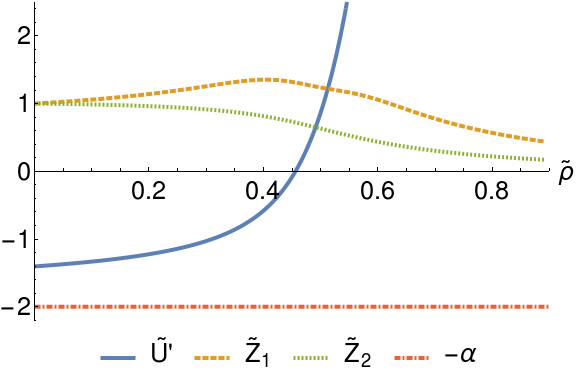}
    \subcaption{Critical FP}
    \end{subfigure}
    \begin{subfigure}{.48 \linewidth}
    \includegraphics[width= \textwidth]{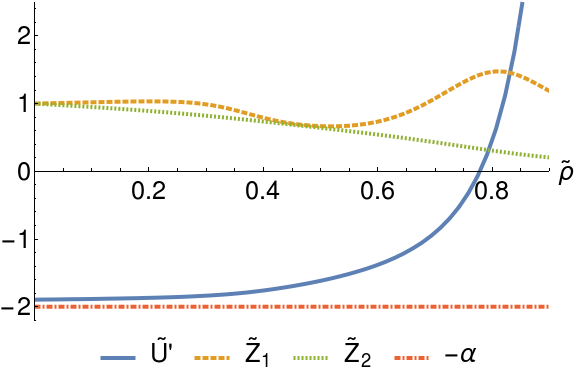}
    \subcaption{QLRO FP}
    \end{subfigure}
    \caption{(Color online) Fixed point functions for $(d,N) = (1.75, 1.3)$. Red horizontal line marks $-\alpha$ corresponding to the propagator pole. Upon decreasing $N$ towards $1$, the fixed point potential builds up singularity as the range of values of $\rho$ where $U'(\rho)$ is close to $-\alpha$ increases.}
    \label{fig:fixed_points}
\end{figure}

In fig. \ref{fig:fixed_points} we juxtapose the functions parametrizing the critical and low-temperature fixed points for $(d,N) = (1.75, 1.3)$. We note two significant qualitative differences between the two FPs. First and arguably more important is the much smaller difference between $U'(0)$ and $-\alpha$ present in the QLRO FP leading to difficulties in the numerical treatment. The second difference is a more complex shape of the $\tilde Z_1$ function featuring two maxima - the first around the minimum of the local potential and the second in significantly lower values of $\tilde \rho$. The interpretation of the structure of the fRG FPs remains somewhat unclear and requires further investigation. We note that on the qualitative level the QLRO FP  is very similar the KT fixed point [$(d,N)=(2,2)$] recovered in the same approximation scheme presented in ref. \cite{Jakubczyk2014} [after a necessary reparametrization].

\begin{figure}
    \centering
    \includegraphics[width=.45 \textwidth]{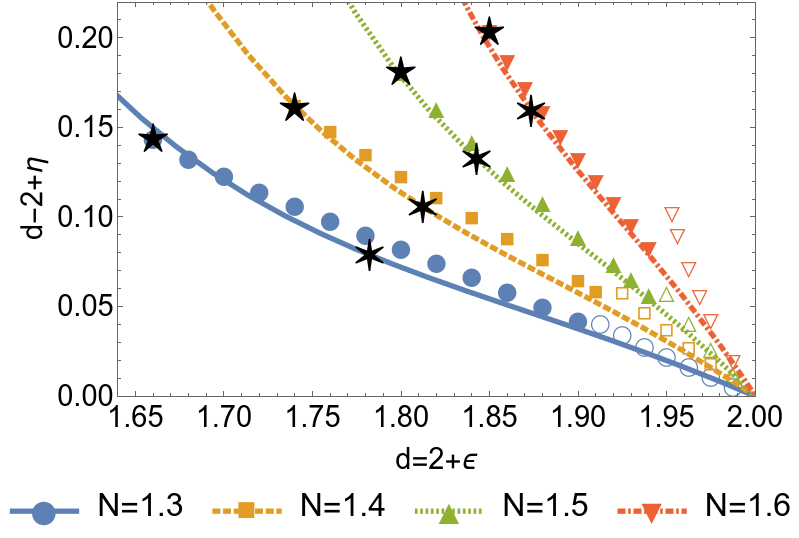}
    \caption{(Color online) Dependence of the correlation function exponent of the QLRO FP on $\epsilon$ for a series of values of $N$. Filled points denote the results from the functional scheme, empty - from the simplified scheme and lines - the predictions of the $2+\epsilon$ expansion [ref. \cite{Bernreuther1986}]. The stars mark the lower critical dimensions: five-pointed as obtained in the functional scheme and six-pointed from the CH scenario. The lines are extended below the point of the FP collision [where the BZJ FP is no longer infrared stable] along the prediction $2+\epsilon$ expansion.}
    \label{fig:eta_lowt}
\end{figure}

In two dimensions, the QLRO FP coincides with the zero-temperature FP. As a consequence, the dimensionless local potential minimum $\tilde \rho_0$ [see eq. \eqref{phi4}], acting as the inverse temperature, takes an infinite value at this FP. This fact can be used to perform an expansion around $\tilde \rho_0=\infty$ similar to that of ref. \cite{greater} but extended to $\epsilon = d-2<0$. With such an expansion we can show that to the first order in $\epsilon$ the $2+\epsilon$ expansion and the NPRG yield the exact same value of the anomalous dimension, regardless of the employed infrared regulator. However, this is no longer true at higher orders. 
It means that the simplified scheme is the most accurate exactly in the dimensions where the functional scheme becomes the most numerically demanding. This offers a unique opportunity to supplement the functional results with the simplified ones. Fig. \ref{fig:eta_lowt} presents the correlation function exponent $d-2+\eta$ of the QLRO fixed point as a function of $d$ and $N$. In the figure, the combination of functional and simplified results is compared with the predictions the $2+\epsilon$ expansion \cite{Bernreuther1986}. We note a very strong agreement between the $2+\epsilon$-expansion and the functional results in all dimensions. The predictions of the simplified scheme, on the other hand, agree with the $2+\epsilon$-expansion only in a direct vicinity of $d=2$ and deviate from it pretty quickly as the dimension decreases.

It is tempting, at this point, to reach out for a comparison with better understood limiting cases such as the Mermin-Wagner line ($d=2$) and the Ising universality class ($N=1$). The comparison with the former is straightforward. The low-temperature FPs of the two-dimensional $O(N)$ models for $2>N>1$ are characterised by vanishing anomalous dimension. This fact is captured exactly within the $2+\epsilon$ expansion and as consequence within the derivative expansion at order $O(\nabla^2)$. The proper way to compare our results to the Ising universality class is much more unclear. Due to lack of the transverse mode (or at least its fraction) leads to a completely different physics. On a more technical level, one can no longer use $Z_2(\rho_\eta)$ as the order-parameter renormalization factor $Z_k$ since $Z_2$ has no physical interpretation for $N=1$. For those reasons, we refrain from addressing the case of $N= 1$ within this study.

\begin{figure}
    \centering
    \includegraphics[width=.45 \textwidth]{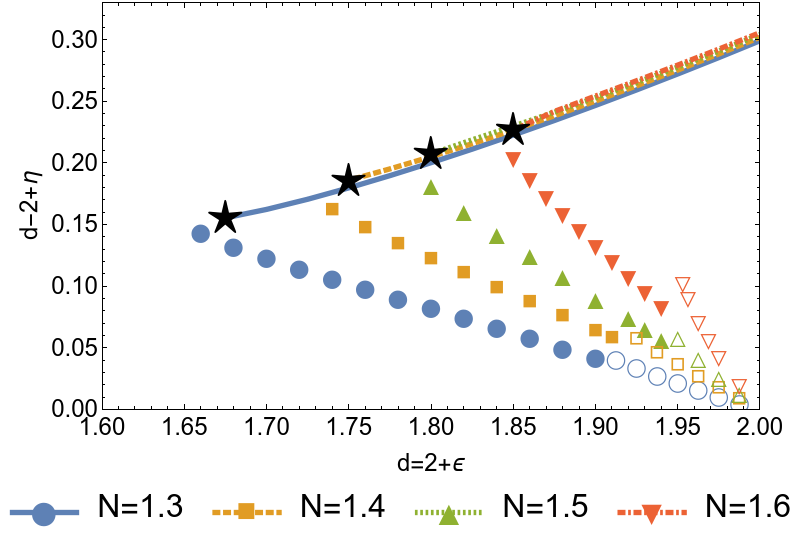}
    \caption{(Color online) The correlation function exponent $d-2+\eta$ of the critical and QLRO FPs as functions of $\epsilon$ for a series of values of $N$. 
    Points denote the exponent of the QLRO FP: filled - from the functional scheme, empty - from the simplified, and lines denote the exponent of the critical FP. Stars mark the lower critical dimensions from the analysis of the critical FPs in the functional scheme.
    }
    \label{fig:eta_my}
\end{figure}

Fig. \ref{fig:eta_my} illustrates the collision of the QLRO and the critical FPs by presenting the evolution of the correlation function exponent $d-2+\eta$ with the changing dimension. The figure uses the data from the functional calculation  wherever available and supplements it with the data from the simplified in dimension close to $d=2$. The collision seems very sharp when compared to fig. \ref{fig:eta_joint}, where it occur almost smoothly. We note a slight discrepancy between the lower critical dimension coming from the analyses of the critical and the QLRO FPs arising due to limited numerical precision.

\subsection{Lower critical dimension}
\begin{figure}
    \centering
    \includegraphics[width=.45 \textwidth]{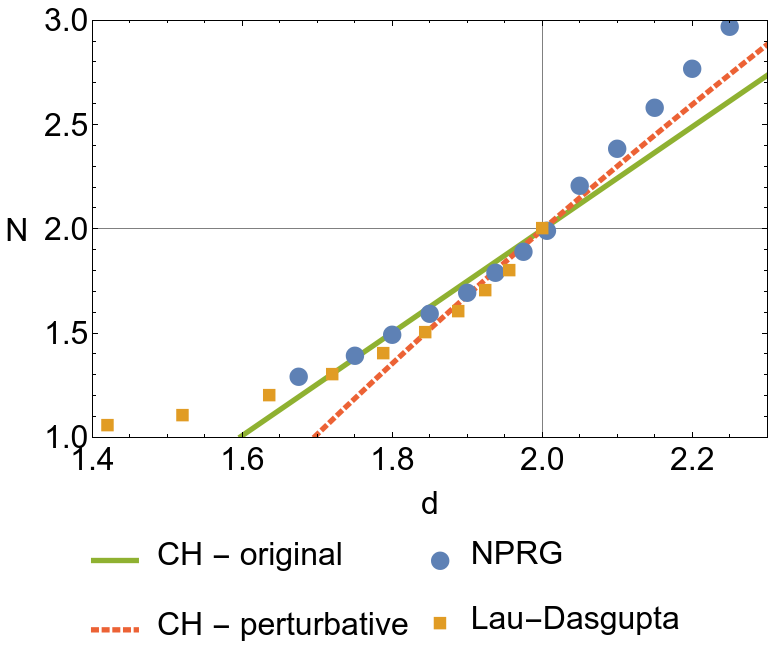}
    \caption{(Color online) Comparison of different estimates of the shape of the line of the lower critical dimension. Lines denote two different ways of charting the CH line based on the perturbative RG calculations. Points denote results from the present work and from ref. \cite{Lau1987}. The NPRG points for $d>2$ mark the position of the CH line estimated within the functional RG scheme in ref. \cite{Chlebicki2021}, where it was obtained in a form of a crossover. }
    \label{fig:ch_line}
\end{figure}

We finally examine the line of the lower critical dimension $d_c(N)$. We estimate $d_c(N)$ by the lowest dimension at which we are able to extract the PMS value for the dominant RG eigenvalue in the functional calculation for a given value of $N$. In fig. \ref{fig:ch_line}, we compare our results with the real space RG data from ref. \cite{Lau1987} as well as two distinct estimates springing from the CH scenario. We recall that the CH line is defined by the equation
\begin{equation}
0 = \Delta = \epsilon \frac{\pi}{2} - (N-2) f\left(\frac{\pi}{2},N\right) +O(\epsilon^2). \label{delta2}
\end{equation}
The first estimate for the CH line follows the original reasoning of Cardy and Hamber which is based on an additional postulate that $f(g=\pi/2,N=2) = 2/\pi$ \footnote{This postulate was introduced to satisfy the analytic form for the first RG eigenvalue conjectured by the authors.}. In the alternative estimation, we employ the perturbative expansion of $f(g,N)$ presented in eq. \eqref{f_perturbative}.

Our results fit very nicely into the presented picture, laying very close to the predictions of refs. \cite{Cardy1980,Lau1987}. It is instructive to compare the different estimates of a slope of the CH line at the KT point $(d,N)=(2,2)$. This data is presented in table \ref{tab:ch_slope}. Our results lay in between the predictions of other works. We note at this point that the perturbative CH prediction varies strongly with the employed order of $g$ expansion and can be not yet sufficiently converged at the present order.


\begin{table}
    \centering
    \begin{tabular}{lr}\toprule
        Calculation & $\left.\frac{\partial N_c(d)}{\partial d}\right|_{d=2}$  \\ 
        \hline CH - original & $\frac{\pi^2}{4} \approx 2.5$\\
        CH - perturbative & $3.1$\\
        NPRG & $4.3$\\
        Ref. \cite{Lau1987} & $5.8$ \\ \toprule
    \end{tabular}
    \caption{Different estimates of the slope of the CH line at the KT point $(d,N)=(2,2)$.}
    \label{tab:ch_slope}
\end{table}

\section{Conclusion}
In this paper we addressed the $O(N)$ models for $d<2$ and $N<2$. We emphasized a consequence of the CH type analysis that below $d=2$ a stable finite-temperature FP emerges from the zero-temperature FP. The existence of the low-temperature FP destabilizes the long-range order and leads to formation of a QLRO phase similar to the KT phase but characterized by a universal (temperature independent) anomalous dimension $\eta$. We then demonstrated that the line of collisions of the critical and low-temperature FPs marks the line of the lower critical dimensions $d_c(N)$ of the $O(N)$ models with $1<N<2$ and $d<2$, and bounds the region where the QLRO phase exists in the model's phase diagram.

Subsequently, we employed the nonperturbative RG and obtained a picture supporting the CH scenario for $d<2$ and $N<2$. We verified the existence of the critical and low-temperature FPs below $d<2$ and demonstrated their collision by tracking the evolution of their anomalous dimensions.
We have shown a very strong agreement between the predictions of NPRG and the $2+\epsilon$ expansion for the anomalous dimension of the QLRO FP for all dimensions between $d=2$ and the lower critical dimension $d_c(N)$.

We also examined the line of lower critical dimensions $d_c(N)$. We obtained a shape very similar to the one obtained within the real space RG \cite{Lau1987} with one significant difference. Compared to ref. \cite{Lau1987}, the slope of our estimate at the KT point $(d,N)=(2,2)$ is substantially closer to the prediction of the CH scenario.

It is striking that, while the results obtained for $d<2$ from the NPRG method in the present work are fully compatible with the predictions of the Cardy-Hamber approach, they differ rather fundamentally in the case $d>2$ described in our previous paper \cite{Chlebicki2021}. There, we have found no indication of nonanalyticities of the critical exponents consistent with the CH picture of the FP collision. Instead, we identified a smooth crossover between two regions of the $(d,N)$ plane well characterized in the CH paper.
 On the other hand, in the present work, we clearly recover a sharp FP collision for $d<2$. We also observe that the line $N_c(d)$ obtained here merges smoothly at the KT point $(d,N)=(2,2)$ with the line of cross-overs of the critical exponents presented in ref. \cite{Chlebicki2021}.   

\section*{Acknowledgments}{
We thank Gilles Tarjus for pointing this problem to us as well as reading the initial version of the manuscript and providing useful comments. We are grateful to Bertrand Delamotte for collaboration on related topics and sharing his experience with us and to Nicol\'as Wschebor for reading and commenting on the initial version of the manuscript. We thank Johan Henriksson for useful correspondence.

We acknowledge funding from the Polish National Science Center via grant 2017/26/E/ST3/00211. AC is grateful for the hospitality of Instituto de F\'isica, Facultad de Ingenier\'ia, Universidad de la Rep\'ublica, Montevideo, Uruguay, where a substantial part of the paper was completed as part of the project BPN/BEK/2021/1/00293 funded by Polish National Agency for Academic Exchange NAWA.
}

\section*{Appendix}
In this section we present the RG equations that were used in this work. To simplify the expressions we denote the `dressed' propagators as follows:
\begin{align}
&G_1(\rho) = \left(Z_1(\rho)q^2 + U'(\rho)+2\rho U''(\rho) + R(k, \bm q^2)\right)^{-1},\\
&G_2(\rho) = \left(Z_2(\rho)q^2 + U'(\rho) + R(k, \bm q^2)\right)^{-1},
\end{align}
In the equations below we use $\int_q$ to denote $\int \frac{d^dq}{\left(2\pi\right)^d}$ and we suppress the $k$ dependence to simplify the notation. The $\beta$ functions employed in the functional calculation read:
 \begin{widetext} 
\begin{align}
    \beta_{U'}(\rho) &=k \partial_k U'(\rho) = -\frac{1}{2} \int_q \left(k\partial_k R\left(q^2\right)\right) \left\{G_1(\rho ){}^2 \left(q^2 \left(Z_1'(\rho )\right)+2 \rho  U^{(3)}(\rho )+3 U''(\rho )\right)+(N-1) G_2(\rho ){}^2 \left(q^2 Z_2'(\rho )+U''(\rho )\right)\right\}, \\
    \beta_{Z_1}(\rho) &= k\partial_k Z_1(\rho) = -\frac{1}{2d}\int_q \left(k\partial_k R\left(q^2\right)\right) \Bigg\{ 4 (N-1) \rho  G_2(\rho ){}^4 \left(q^2 Z_2'(\rho )+U''(\rho )\right) \Bigg[-\left(\left(\partial_{q^2} R\left(q^2\right)\right)+Z_2(\rho )\right) \left((d+4) q^2 Z_2'(\rho )+d U''(\rho )\right)\nonumber \\
    &-2 q^2 \left(\partial_{q^2}^2 R\left(q^2\right)\right) \left(q^2 Z_2'(\rho )+U''(\rho )\right)\Bigg]+4 (N-1) G_2(\rho ){}^3 \left(d \left(Z_1(\rho )-Z_2(\rho )\right) \left(q^2 Z_2'(\rho )+U''(\rho )\right)+q^2 \rho  Z_2'(\rho ){}^2\right) \nonumber \\
    &+4 \rho  G_1(\rho ){}^4 \left(q^2 Z_1'(\rho )+2 \rho  U^{(3)}(\rho )+3 U''(\rho )\right) \Bigg[-\left(\left(\partial_{q^2} R\left(q^2\right)\right)+Z_1(\rho )\right) \left((d+4) q^2 Z_1'(\rho )+d \left(2 \rho  U^{(3)}(\rho )+3 U''(\rho )\right)\right) \nonumber \\
    &-2 q^2 \left(\partial_{q^2}^2 R\left(q^2\right)\right) \left(q^2 Z_1'(\rho )+2 \rho  U^{(3)}(\rho )+3 U''(\rho )\right)\Bigg]+4 \rho  G_1(\rho ){}^3 Z_1'(\rho ) \left((2 d+1) q^2 Z_1'(\rho )+4 d \rho  U^{(3)}(\rho )+6 d U''(\rho )\right) \nonumber \\
    &-d G_1(\rho ){}^2 \left(Z_1'(\rho )+2 \rho  Z_1''(\rho )\right)+16 (N-1) q^2 \rho  G_2(\rho ){}^5 \left(\left(\partial_{q^2} R\left(q^2\right)\right)+Z_2(\rho )\right){}^2 \left(q^2 Z_2'(\rho )+U''(\rho )\right){}^2  \nonumber \\
    &+16 q^2 \rho  G_1(\rho ){}^5 \left(\left(\partial_{q^2} R\left(q^2\right)\right)+Z_1(\rho )\right){}^2 \left(q^2 Z_1'(\rho )+2 \rho  U^{(3)}(\rho )+3 U''(\rho )\right){}^2\nonumber \\
    &-\frac{d (N-1) G_2(\rho ){}^2 \left(\rho  Z_1'(\rho )-Z_1(\rho )+Z_2(\rho )\right)}{\rho }\Bigg\}, \\
    \beta_{Z_2}(\rho) &= k\partial_k Z_2(\rho) = -\frac{1}{2d \rho}\int_q \left(k\partial_k R\left(q^2\right)\right)\Bigg\{G_2(\rho ){}^2 \Bigg[-d \left((N-1) \rho  Z_2'(\rho )+2 \left(\partial_{q^2} R\left(q^2\right)\right)+2 Z_2(\rho )\right) \nonumber \\
    &-4 q^2 G_1(\rho ) \left(\left(\partial_{q^2} R\left(q^2\right)\right)-\rho  Z_2'(\rho )+Z_2(\rho )\right) \left(\left(\partial_{q^2} R\left(q^2\right)\right)+\rho  Z_2'(\rho )+Z_2(\rho )\right)-4 q^2 \left(\partial_{q^2}^2 R\left(q^2\right)\right)\Bigg] \nonumber \\
    &+G_1(\rho ){}^2 \left(-d \left(2 \left(\partial_{q^2} R\left(q^2\right)\right)+2 \rho ^2 Z_2''(\rho )+5 \rho  Z_2'(\rho )+Z_1(\rho )+Z_2(\rho )\right)-4 q^2 \left(\partial_{q^2}^2 R\left(q^2\right)\right)\right) \nonumber \\
    &+G_2(\rho ) \Bigg[G_1(\rho ) \left(4 d \left(\left(\partial_{q^2} R\left(q^2\right)\right)+\rho  Z_2'(\rho )+Z_2(\rho )\right)+8 q^2 \left(\partial_{q^2}^2 R\left(q^2\right)\right)\right) \nonumber \\
    &+4 q^2 G_1(\rho ){}^2 \left(\left(\partial_{q^2} R\left(q^2\right)\right)+\rho  Z_2'(\rho )+Z_2(\rho )\right) \left(-\left(\partial_{q^2} R\left(q^2\right)\right)+\rho  Z_2'(\rho )-2 Z_1(\rho )+Z_2(\rho )\right)\Bigg] \nonumber \\
    &+4 q^2 G_1(\rho ){}^3 \left(\left(\partial_{q^2} R\left(q^2\right)\right)+Z_1(\rho )\right){}^2+4 q^2 G_2(\rho ){}^3 \left(\left(\partial_{q^2} R\left(q^2\right)\right)+Z_2(\rho )\right){}^2\Bigg\}.
\end{align}
\end{widetext}

\bibliography{bibliography.bib}
\bibliographystyle{apsrev4-1}
\end{document}